\documentclass{emulateapj}
\usepackage{natbib}
\usepackage{graphicx,amssymb,amsmath}
\usepackage[dvipsnames]{color}

\shorttitle{Energy deposition and entropy in REXCESS clusters}
\shortauthors{Chaudhuri, Nath, Majumdar}

\begin{document}

%%%%%%%%%%%%%%%%%%%%%%%%%%%%%%%%%%%%%%%%%%%%%%%%%%%%%%%%%%%%%%%%%%%%%%%%%
\newcommand{\half}{\frac{1}{2}}
\newcommand{\3}{\ss}
\newcommand{\n}{\noindent}
\newcommand{\eps}{\varepsilon}
\def\be{\begin{equation}}
\def\ee{\end{equation}}
\def\ba{\begin{eqnarray}}
\def\ea{\end{eqnarray}}
\def\de{\partial}
\def\msun{M_\odot}
\def\div{\nabla\cdot}
\def\grad{\nabla}
\def\rot{\nabla\times}
\def\ltsima{$\; \buildrel < \over \sim \;$}
\def\simlt{\lower.5ex\hbox{\ltsima}}
\def\gtsima{$\; \buildrel > \over \sim \;$}
\def\simgt{\lower.5ex\hbox{\gtsima}}
\def\etal{{et al.\ }}
\def\red{\textcolor{red}} 
\def\blue{\textcolor{blue}}
\def\del{\partial}
\newcommand{\pd}{\partial}

%%%%%%%%%%%%%%%%%%%%%%%%%%%%%%%%%%%%%%%%%%%%%%%%%%%%%%%%%%%%%%%%%%%%%%%%%

\title{Energy deposition profiles and entropy in galaxy clusters
%\\Energy deposition and entropy profiles in nearby galaxy clusters (REXCESS)
}

\author{Anya Chaudhuri$^1$, Biman B. Nath$^2$, Subhabrata Majumdar$^1$}
\affil{$^1$Tata Institute of Fundamental Research, 1, Homi Bhabha Road, Mumbai 400005, India}
\affil{$^2$Raman Research Institute, Sadashiva Nagar, Bangalore 560080, India}
\email{anya@tifr.res.in, biman@rri.res.in, subha@tifr.res.in}

\begin{abstract}
We report the results of our study of fractional
entropy enhancement in the intracluster medium (ICM) of the clusters
from the representative
{\it XMM-Newton} cluster structure survey (REXCESS).
We compare the observed entropy profile of these clusters with that
expected for the ICM without any feedback, as well as with the introduction
of preheating and cooling. We make 
the first  estimate of the total, as well as radial, non-gravitational energy deposition  upto $r_{500}$
for this large, nearly flux-limited, sample of clusters.  
We find that the total
energy deposition corresponding to the entropy enhancement is proportional
to the cluster temperature (and hence cluster mass). The energy deposition, scaled by $T_{sp}$, per particle
as a function of gas mass shows a similar profile in all clusters, with
its being more pronounced in the central region than in the outer region.
Our results support models of entropy enhancement
through AGN feedback.
\end{abstract}


\section{Introduction}
Models of structure formation in the universe have been successful in 
predicting the average properties of galaxy clusters. Using these 
characteristics, such as gas temperature, XRay luminosity, SZ-flux and  richness, it is
possible to draw cosmological conclusions from surveys of galaxy clusters (e.g., \cite{reiprich02,
vikhlinin09, gladders07, khedekar10, rozo10, sehgal11, benson11}.
The detailed properties of the intracluster medium (ICM)  however, need more
input to the physics of baryonic gas than its falling into a dark matter
potential (e.g. \cite{shaw10, battaglia11, trac11, chaudhuri11}). It is believed that feedback from galaxies, including active
galactic nuclei (AGNs), and/or radiative cooling of the ICM gas, modify the
X-ray properties of the gas (see \cite{mcnamara07,mcnamara12}). These
non-gravitational processes tend to increase the entropy of the ICM gas,
thereby making it tenuous, and consequently, under-luminous in X-rays, 
especially in low temperature (and mass) clusters.

Recent observations of profiles of entropy (defined as $K=k_BT/n_e^{2/3}$, where
$n_e$ is the electron number density and $k_B$, the Boltzmann constant \footnote{We write $K$ for the entropy
popularly defined in X-ray literature, and denote the thermodynamic entropy as $S$})
allow one to compare them with
theoretically expected profiles with or without feedback, and allow one
to determine the nature and degree of feedback. Entropy as defined above
is well suited for this sort of analysis as it is a record
of the accretion of gas into the cluster, as well as  the modifications
shaped by the processes of gas cooling and feedback.  \cite{voit05} had
shown that in the absence of any feedback and cooling processes, simulations
tend to predict a power-law radial profile for the entropy outside the core, with a scaling
$K \propto r^{1.1}$. 

Since entropy per particle is a Lagrangian quantity, it is more sensible to 
study the distribution of entropy not with the radial distance, but with
gas mass, taking into account the movement of gas shells due
to a change in entropy.
\cite{voit05} suggested the comparison of entropy as a function of gas mass
in order to determine the enhancement of entropy from non-gravitational 
processes (see also \cite{nath11}). Here, we study the entropy
profiles of clusters from the {\it REXCESS} sample 
and compare with the baseline profiles
of ICM without any feedback. For these clusters, \cite{pratt10} studied the radial entropy profiles
and after comparing with the initial profile, found that entropy enhancement
is evident in the inner radii, and that it extends up to a large radii for low mass systems,  while large mass clusters do not show
entropy deviation at very large radii. In this {\it Letter}, we focus on the
 entropy profiles as functions of gas mass contained inside a given shell. 
 We determine $TdK/K$ , where $dK/K$ is the fractional deviation
of the observed entropy from the benchmark theoretically calculated entropy,
which is a measure of the energy deposition per unit gas particle, and investigate the
profile of this energy deposition for low and high temperature clusters. 

We adopt a $\Lambda$CDM cosmology with $H_0=70$ km s$^{-1}$ Mpc$^{-1}$,
$\Omega_M=0.3$ and $\Omega_\Lambda =0.7$.

\section{The cluster sample}
The REXCESS survey \citep{bohringer07} uses the REFLEX cluster catalog as a
parent sample. REFLEX
is a nearly complete flux limited cluster sample, covering 4.24 ster in the
southern
extragalactic sky \citep{bohringer04}. This sample consists of 31
local clusters in
the redshift
range $z\leqslant 0.2$. The clusters are selected on the basis of their
X-Ray luminosity, $L_X = (0.407 \hbox{--}20) \times 10^{44}$ $h_{50}^{-2}$ erg
$s^{-1}$ in the $0.1\hbox{--} 2.4$ keV
band, with a homogeneous coverage in the chosen luminosity range, and no
preference for any morphology type.
The selected luminosity range provides clusters with a temperature $\simgt2$ keV,
 and does not include
galaxy groups. As \cite{pratt10} have noted, the properties of the REXCESS sample
allow one to study the variation of entropy profiles across a range of cluster
masses, especially because the distances were chosen such that $r_{500}$ fell
within the {\it XMM-Newton} field of view, which increased the precision of
measurements at large radii. They also subdivided
the sample into cool-core and non cool-core systems, defining the clusters
with central density $E(z)^{-2}n_{e,0} > 4 \times 10^{-2}$
cm$^{-3}$ as cool-core systems ($E(z)$ being the ratio of the Hubble constant at
redshift $z$ to its present value).

%\cite{croston08} obtained the radial density profiles from the surface brightness
%profiles of the REXCESS sample,
%centred on the peak of the X-ray emission, in the $.3\hbox{--}2$ keV
%band. \cite{pratt10} have studied the radial entropy profile, and also 

%Since the density profiles are determined on a radial grid
%of significantly higher resolution than that of the temperature
%profiles, \cite{pratt10} determined the best fitting parametric 3D
%temperature profile on the same grid as that of the deprojected, deconvolved density profile, and calculated the entropy ,
%$K = kT/n_e^{2/3}$.

In this work, we use the entropy profiles of 25 clusters from the whole 
REXCESS sample of 31 clusters (see \cite{pratt10}, their Table1). We use only those clusters with data at 
a minimum of 5 radial points outside the core radii, thus excluding clusters
number 2, 13, 23, 25 \& 27 (ordered top-to-down respectively in the table).  We also leave out cluster number 14 whose errors on observed entropy far exceed those of other clusters.

\section{Entropy profiles}
\subsection{Initial entropy - radial profile}
In order to assess the entropy enhancement in observed clusters, we first
discuss the profile expected without any non-gravitational processes.
\cite{voit05} presented an analytic form for the baseline entropy profile 
which they obtained
by analyzing the entropy profiles of clusters  from 
non-radiative simulations.
Their simulated SPH profiles, when fitted
in the $0.1\hbox{--}1$ $r_{200}$ range, scatter about a median scaled profile
described by a baseline power law relation,
\begin{equation}
 \frac{K (r)}{K_{200}} = 1.32   (\frac{r}{r_{200}})^{1.1}
\label{baseline}
\end{equation}
 with approximately $20 \%$ dispersion. These profiles however are found to
flatten within a radius of $R < 0.2 $ $r_{500}$, and they find that the agreement of the 
above fit with both their SPH and AMR simulations is better than $\sim 10\%$ beyond a radius
of $0.1$ $r_{200}$.

%\begin{centering}
\begin{figure}[h]
\hspace*{-0.5cm}
\includegraphics[scale=.51]{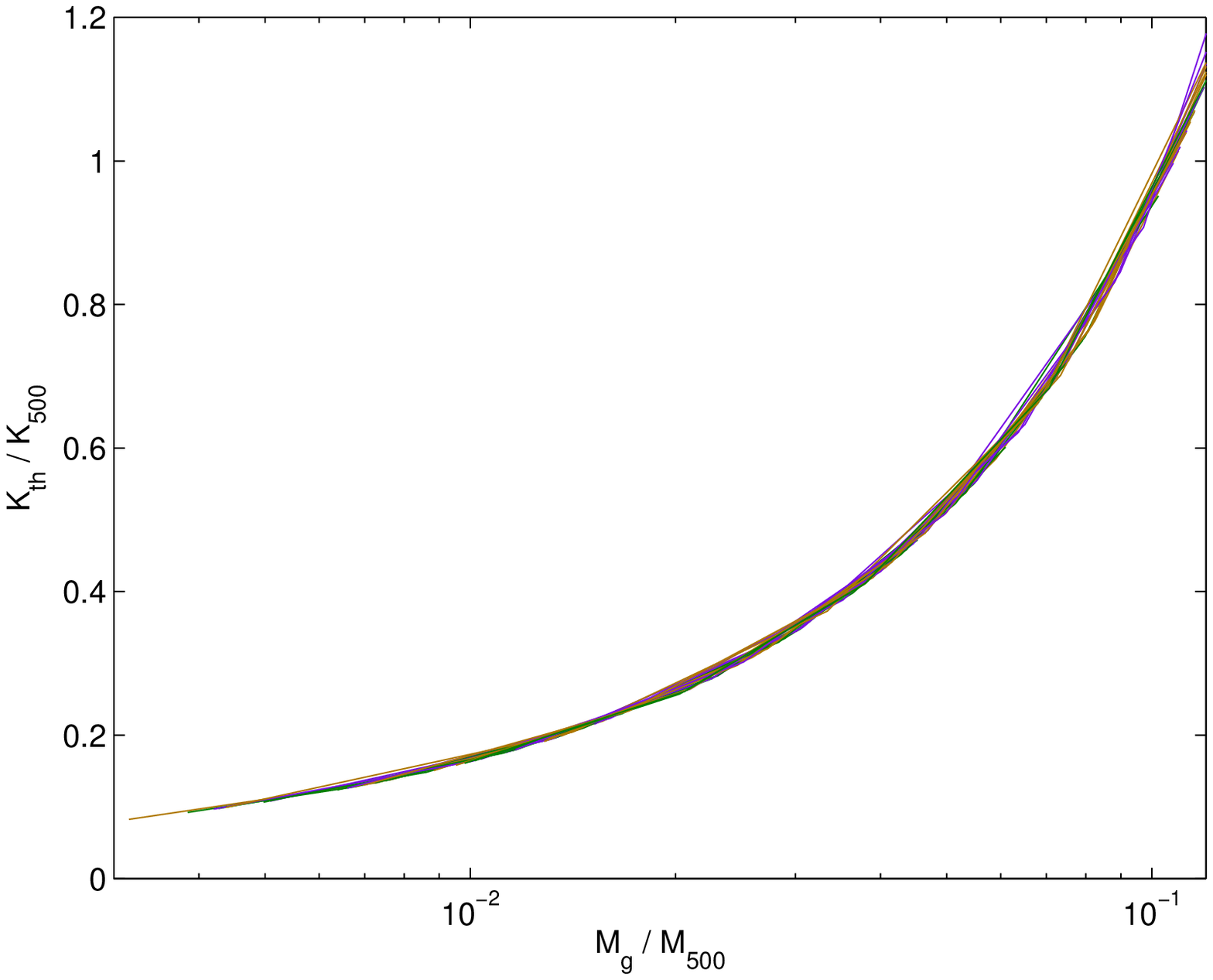}
\includegraphics[scale=.45]{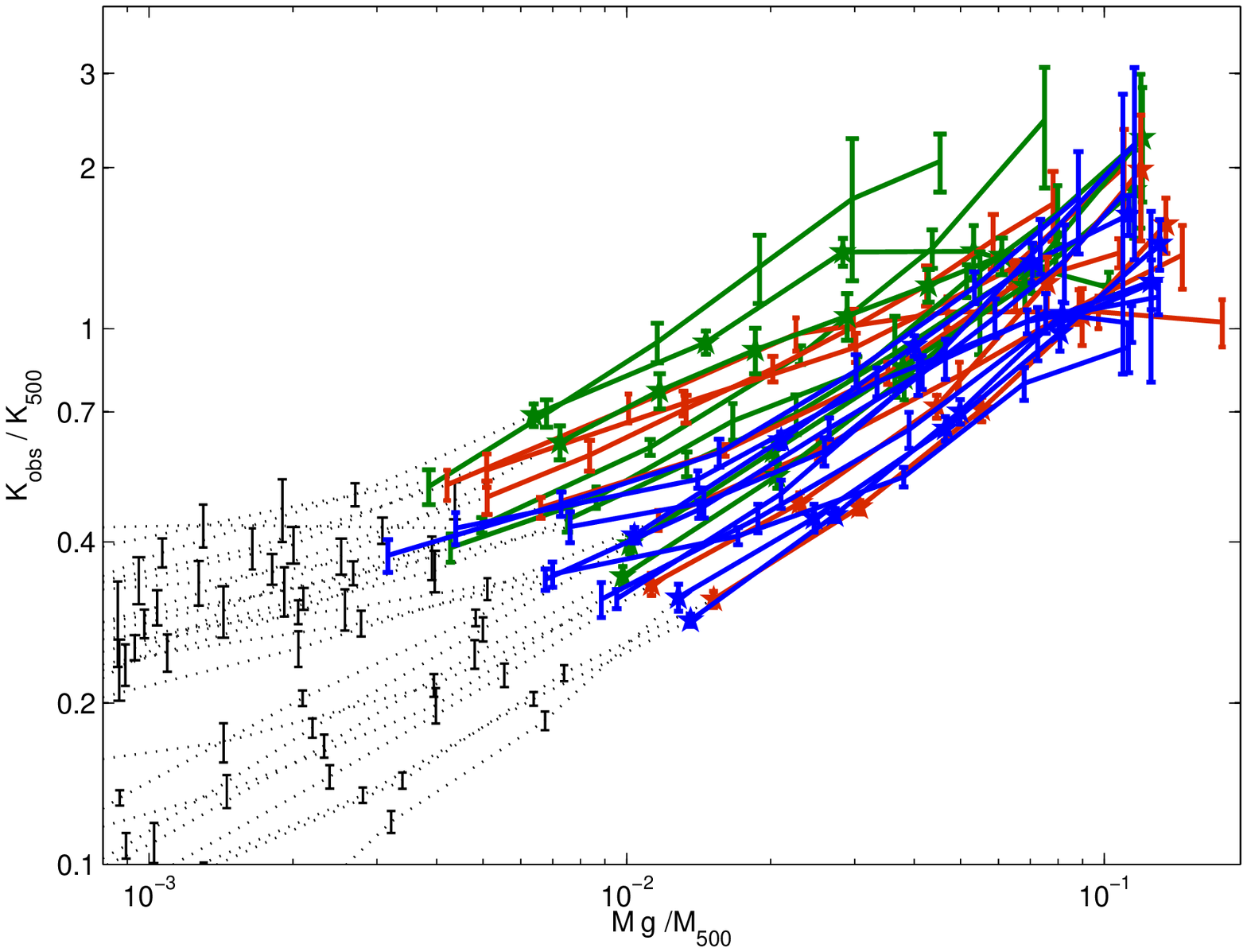}
\caption{
Upper Panel - This plot shows the ratio of $K_{th}/K_{500}$ as a function of gas
mass $M_g$ for all the clusters. 
Green lines refer
to the lowest temperature clusters ($T_{sp}\le 3.5$ keV), 
red for the intermediate temperature clusters 
($3.5 \le T_{sp}\le 5$ keV), and blue lines are for the largest 
temperature clusters ($T_{sp}\ge 5$ keV). Lower Panel - $K_{obs}/ K_{500}$ is 
plotted against $M_g/M_{500}$. Color scheme is the same as above. Dotted lines show
observations below $0.1r_{200}$. $T_{sp}$ is the mean spectroscopic temperature
in the $0.15 $ - $0.75$ $r_{500}$ range \citep{pratt10}.
}
\label{k_by_k500}
\end{figure}
%\end{centering}

\subsection{Initial entropy profile with gas mass}
In this paper, we would like to study the entropy as a function of gas mass.
In order to calculate the initial entropy profile as a function of gas mass, 
we  use the initial radial entropy profile,
 in conjunction with the assumption of hydrostatic
equilibrium.
 We assume the Navarro-Frenk-White (NFW) profile for the dark matter
halo \citep{nfw97}. For the  concentration parameter  
$c_{500}=3.2$ that we adopt\footnote{This value is measured for a
morphologically relaxed cluster sample by \cite{pointecouteau05}, also used by \cite{pratt10}.}, 
the corresponding relation for eqn  ~\ref{baseline} at $r_{500}$ becomes,
\citep{pratt10}
\begin{equation}
\frac{K( r )}{K_{500}} = 1.42 \, ( \frac{ r }{ r_{500}} )^{1.1}
\label{entbase}
\end{equation}
The equation of hydrostatic equilibrium can be written as,
\begin{equation}
\dfrac{dP_g}{dr} = - \rho_g \dfrac{G M( <r )}{ r^2} 
=-\left[ \dfrac{P_g}{K} \right ]^{3/5} m_p \mu_e^{2/5} \mu^{3/5} \dfrac{ G M( <r ) }{ r^2} 
\label{he2}
\end{equation}
where $P_g = n_g k_B T$, is the gas pressure. 
For boundary conditions, we set the total gas fraction inside $r_{vir}$ to
be the universal baryon fraction, $f_g$ = $\Omega_b/\Omega_m$. 
Eqns \ref{entbase} \& \ref{he2} are solved for the pressure profile $P_g$; this gives the density profile and hence $M_g$. One can then invert $M_g$ to get $K(M_g)$.

We show the results in Figure \ref{k_by_k500} (Upper Panel) that plots the theoretical profiles of $K_{th}$
scaled by $K_{500}$, the characteristic entropy (eq 3 in \cite{pratt10}), for clusters of different
temperature bins.  The figure shows that $K_{th} (M_g)/K_{500}$ is self similar to a good approximation. 
We have fitted the profile with a simple
parameterization, $K_{th} (M_g) /K_{500} =A \,
(M_g/M_{500})^\alpha$, in the range $0.1r_{200} - r_{500}$. 
The scatters in the slopes and normalization  
for different values of $M_{500}$ are found to be small.
The index $\alpha=0.81 \pm 0.05$ 
for the whole sample. The value of the parameter $A$ was determined to be
$6.09 \pm 0.86$.

\subsection{Observed entropy - radial profile}
\cite{pratt10} have fit the REXCESS data to the form,
\begin{equation}
K(r) = K_{0} + K_{100}  [r/100 \; {\rm kpc}]^\alpha
\end{equation}
where $K_{0}$ is interpreted as the excess
of core entropy above the best fitting power law at large radii.
They scaled the quantities to $r_{500}$,
the effective limiting radius for high quality observations
from {\it XMM-Newton}
and {\it Chandra}, which they estimated iteratively from the
updated calibration of the $M_{500}-Y_X$ relation, by including
the REXCESS data for morphologically relaxed systems. 
Interior to $r_{500}$, the observed entropy is always higher than the baseline 
prediction. At $r_{500}$, they find the median dimensionless entropy is $K
(r_{500})/K_{500} = 1.70 \pm .35$ and that this is higher than but 
consistent with the baseline prediction.

\subsection{Observed entropy profiles with gas mass}
Next, we express the observed entropy profiles $K_{obs}/K_{500}$ of REXCESS
clusters as a function of  $M_g/M_{500}$. Figure \ref{k_by_k500} (Lower Panel) shows these profiles 
for all 25 clusters in our sample 
in different temperature ranges respectively. We fit the profiles
by an
expression of the form $K_{obs}/K_{500} = A_o + B_o \, (M_g/M_{500})^{\alpha_o}$,
in the range $0.1r_{200}-r_{500}$. We give mean and {\it rms} of the best fit values in Table \ref{obstable1}.
Note that the power-law indices for the observed entropy-gas mass 
relation are shallower than expected from the theoretical expectation. 
Interestingly this index (logarithmic slope) does not differ much
in the whole cluster sample. 
The values of $A_o$ and $B_o$ show
that cool-core clusters are entropy deficient.

\begin{table}
\caption{Mean values of parameters in the range $0.1r_{200}-r_{500}$ (excluding core): observed entropy-gas mass relation}
\label{obstable1}
\vskip 0.2cm
  \begin{tabular}{|c|c|c|c|}
                  \hline
{\small Sample}&$A_o$&$B_o$&$\alpha_o$ \\
                   \hline
                   \hline
Total sample (25 clusters)& $0.23 \pm 
0.43$& $9.59 \pm 7.54$ & $0.67 \pm 0.47$ \\
                   \hline
cool-core (9)& $0.14 \pm 0.61$  & $9.07 \pm 6.18$ &
$0.63 \pm 0.61$ \\
  \hline
Non cool-core& $0.28 \pm 0.30$&$9.89 \pm 8.38$ &
$0.69 \pm 0.40$ \\
                   \hline
\end{tabular}
\end{table}

\section{Effects of only preheating and cooling}
\cite{voit02} discussed three types of modifications to the initial
entropy profile, namely: (i) a truncation in the entropy profile owing to removal
of gas, approximating the effect of gas cooling and dropping out of the ICM, 
(ii) a shift in the profile, 
mimicking the effect of preheating, and (iii) lowering the entropy profile due
to radiative cooling. Assuming a form of the cooling function of the type
$\Lambda \propto T^{-1/2}$ for group temperatures $(T \le 2$ keV), 
it was shown that $K^{3/2}$ across the cluster
is reduced by an
amount $\frac{3}{2}K_{c}^{3/2}$, where $K_{c}$ is a critical entropy.
\cite{johnson09}
suggested a combination of the effects of preheating
and cooling, expressed as
\begin{equation}
%S_{X,obs} =  \left[  ( S_{X,th} + S_{X,shift} )^{3/2} - A \right]^{2/3} \,
K_{obs}^{3/2} = (K_{th} + K_{shift})^{3/2}-{3 \over 2} K_{c}^{3/2}
\label{fitform}
\end{equation}
$K_{c}  \approx 81 \, {\rm keV cm^2} [T/1 \, 
{\rm keV}]^{2/3} \, [t/14 \, {\rm Gyr}]^{2/3}$, (their eqn 14)
describes the cooling. 
They calculated the constant preheating shift  $K_{shift}$ by evaluating eqn \ref{fitform} at their outermost radial point.

\begin{figure}[h]
% \centering
 \includegraphics[scale=0.45]{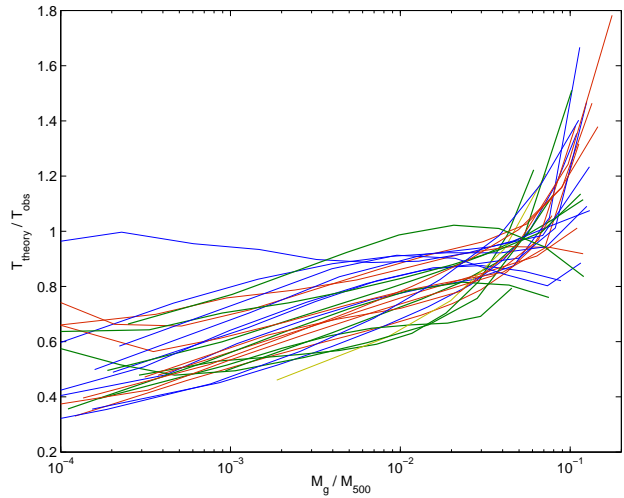}
\caption{
The ratio of the theoretical to observed temperature as a function of 
$M_g/M_{500}$ is plotted for our sample of 25 clusters.
}
\label{ratiotemp}
\end{figure}

For the 28 nearby galaxy groups from the {\it XMM-Newton} survey, \cite{johnson09} 
found that while this 'preheating + cooling' model matches the
observations better than a simple shift/truncation, it still fell short of being
a reasonable representation of the observed profiles.

We have fitted models of the form of eqn \ref{fitform} to our sample, where
we have used the entire profile $K_{th} (M_g)$ and $K_{obs} (M_g)$, rather than just one radial point
for the fit. 
 We attempted three different types of fits for each cluster, described
below:
(i) $K_{C}$ is evaluated using the full radial temperature profile instead of mean temperature;
(ii) A fit using the constant $T_{sp}$ for each cluster; 
(iii) We assume that a fraction of the gas mass is lost from the ICM, and
try two fits with varying fractions of the total gas mass, $f=0.8,\, 0.9$.
The temperature  $T = T_{sp}$, and the expression used for the fitting is:
$K_{obs}( f \, M_g) =  \left[  ( K_{th}(M_g) + K_{shift} )^{3/2} - 
(3/2)K_C^{3/2} \right]^{2/3}$. 

We find that the number of clusters for which none of fits are good far exceeds the clusters for which
any of fits can be called reasonable (reduced $\chi^2 < 2$). 
The lack of a good fit to the preheating+cooling model in most of the clusters
in the sample suggests that a major component of entropy enhancement occurs
beyond simple preheating and radiative cooling.

\section{Fractional entropy deviation and energy input}
In order to determine
the amount of energy deposition associated with the entropy enhancement,
we use the quantity $T_{obs} \Delta K / K_{obs}$,
where $\Delta K =  K_{obs} - K_{th}$. This is because the 
thermodynamic entropy of an ideal gas $S$ is related to $K$, as $S={\rm Const.} \times \ln K$  
and the change in energy per
unit mass $dQ = T dS \propto T \Delta K / K$ 
(see also eqn 3 of \cite{finoguenov08}).

\begin{figure}[h]
%\centering
\includegraphics[scale=0.45]{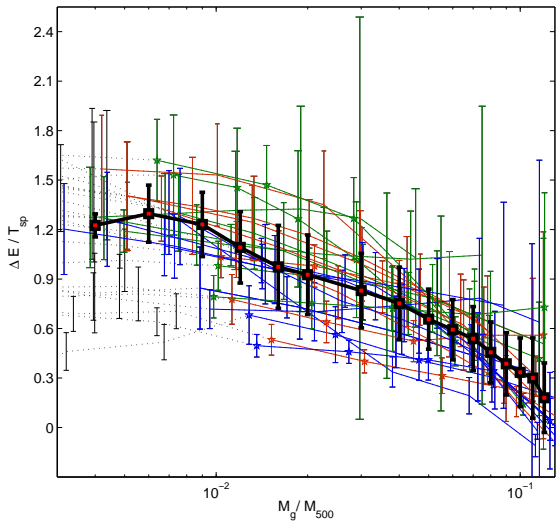}
\caption{
The profiles of energy deposition per unit particle plotted against $M_g/M_{500}$, after scaling them by $T_{sp}$ for different clusters. Data points inside the core radii are shown in dotted lines. 
The color scheme is same as that used in fig \ref{k_by_k500}.
The mean profile and 1-$\sigma$ scatter, outside the core radii, is shown with thick black line.
}
\label{deltaEperpartmass}
\end{figure}

In  an isochoric process, (see also \citep{lloyd-davies00}),
\begin{equation}
\Delta Q = {\Delta K  n_e^{2/3} \over (\gamma -1) }= {kT_{obs} \over
(\gamma-1) } {\Delta K \over K_{obs}} \,. 
\end{equation}
In  an isobaric process, however,  (for $T_f/T_i=\beta$)
\begin{eqnarray}
\Delta Q &=& {\Delta K  n_f^{2/3} \over (1-{1 \over \gamma}) \mu m_p}
{(\beta^{5/3}-1) \over \beta ^{2/3} (\beta -1)}\\ \nonumber
&=& {kT_{obs}  \over (1-{1 \over \gamma}) \mu m_p} 
{(\beta^{5/3}-1) \over \beta ^{2/3} (\beta -1)}
{\Delta K \over K_{obs}} \,. 
\end{eqnarray}
The ratio of the changes in energy for a given fractional change  
$\Delta K/K_{obs}$ and $T_{obs}$, is given by,
\begin{equation}
{\Delta Q _{isobaric} \over \Delta Q _{isochoric}}
=\gamma {(\beta^{5/3}-1) \over \beta ^{2/3} (\beta -1)} \,.
\end{equation}
For a value of $\beta=2$, the ratio is $1.14$. This implies that if the
observed temperature $T_{obs}(M_g)$ deviates from the theoretically calculated value $T_{th} (M_g)$
by a factor $\le 2$, then the two above mentioned estimates of energy
input per unit mass differ by only a factor of $1.14$. 
Figure \ref{ratiotemp} shows the ratio  $\frac{T_{th}(M_g)}{T_{obs}(M_g)}$  for clusters in the sample. One can see that the two temperature profiles  vary within a factor of $\sim 2$ ; hence,
 we choose the expression for the isochoric process in our estimates.

We first estimate the energy per particle, $\Delta E (M_g)\,= \frac{3}{2}\, T_{obs} {\Delta K \over K_{obs}}$, for each cluster.
Fig \ref{deltaEperpartmass} shows the profiles for $\Delta E/T_{sp}$ , the ratio of the non-gravitational energy injection to the gravitational 
potential of the clusters in three temperature bins. 
  While the
detailed profiles differ from cluster to cluster, $\Delta E/T_{sp}$ generally has a decreasing profile with a similar trend.
Albeit with a large scatter within each temperature group, we find that the mean value of $\Delta E/T_{sp}$ is the highest for 
the lowest temperature group and vice versa. For all clusters, non-gravitational energy is already comparable to gravitational 
energy at core radii. Moreover, 
the profiles decrease by a factor of $\sim50$\% for  $T_{sp} \le 3.5$ keV clusters and $\sim75$\% for $T_{sp} > 3.5$ keV clusters. Thus, on average, the  profiles for higher masses  decrease faster 
than those for lower masses. Our calculations, as mentioned earlier, 
are valid outside the core $0.1 r_{200}$.

We determine the mean profile after averaging over all the $\Delta E$  profile fits for the individual clusters.\footnote{To this end, we use the fit $\Delta E = C + D \, (M_g/M_{500})^\delta$}
The mean profile with the $1-\sigma$ scatter is shown in fig \ref{deltaEperpartmass} .
The mean profile decreases by roughly a factor of 4 between $0.1r_{200}$ and $r_{500}$.

\begin{figure}[h]
\includegraphics[scale=.45]{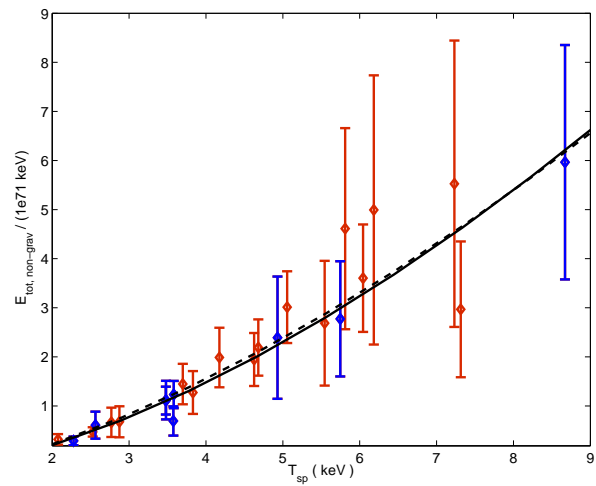}
\caption{
The total energy injection between $0.1r_{200}-r_{500}$ is plotted against cluster mass.
Red points are for non cool-core clusters and blue points show cool-core 
clusters. The best fit to entire sample is shown in solid black line. The black dashed line shows best fit to the non cool-core clusters.
}
\label{deltaEmass}
\end{figure}

The total amount of energy deposited, for the whole cluster is ,  
\begin{equation}
 E_{non-grav} \,= \,\int {kT_{obs} \over (\gamma -1) \mu m_p} {\Delta K \over 
K_{obs}}  \, dM_g \,,
\end{equation}
for $M_g/M_{500}$ between the limits $0.1r_{200}<r<r_{500}$.
We use the fits obtained above to calculate the integral and
the results are shown in Figure \ref{deltaEmass}. Clearly $E_{non-grav}$ is proportional to the cluster mass. A fit results in the following scaling relations for the whole sample:
\begin{equation}
\frac{E_{non-grav}}{10^{71} \, {\rm keV}} = (-0.414 \pm 0.41) + (0.2 \pm 0.17) \, (\frac{T_{sp}}{{\rm keV}})^{1.62 \pm 0.47}.  
\end{equation}
If the cool-core clusters are omitted, then one obtains a slope of
$1.53 \pm 0.64$, a constant term of $-0.48 \pm 0.65$ and first coefficient as $0.24 \pm 0.3$.

Dividing
the energy by the total number of particles in the ICM, we estimate the  
mean energy to be $2.74 \pm 0.87 $ keV per particle.

\section{Discussions}
We can compare
the value of  $\sim 2.74$ keV per particle as a measure of the energy input 
corresponding to the observed entropy enhancement, with those inferred from 
other considerations. 
We note that the earlier observations of entropy enhancement in the ICM,
mostly inferred from the deviations of several cluster scaling relations, were
interpreted in terms of an entropy floor. It was thought that elevating
the ICM gas to an entropy floor of a few hundred keV cm$^2$ would explain the
observations. This entropy floor can be associated with an amount of energy
if the density of the gas at the epoch of energy input is known. For example,
 simulations by \cite{borgani01} showed that an entropy floor of $\sim 50$
keV cm$^2$ was adequate to explain the observations. They found this corresponds
to $\sim 1$ keV per particle if the heating was assumed to have taken place
at $z\sim 3$. 

Interestingly, \cite{roychowdhury04} showed  for their
model of AGN feedback from black holes that  X-ray observations could be explained
with an  energy input  proportional to cluster mass.
In their model of AGN feedback through buoyant bubbles of relativistic plasma,
which deposit energy into the ICM through $pdV$ work, convection and thermal
conduction, this proportionality implied a linear relation between the black
hole mass of the central AGN and the cluster mass. We find that in order
to explain the correlation in Figure \ref{deltaEmass}, we need the black hole mass
$M_{bh} \sim 2 \times 10^{-6} M_{500} \; \eta_{0.2}$, where the energy available
from the AGN is characterized by an efficiency $\eta=0.2$.

Simulations by \cite{gaspari12} also show that the energy 
deposition is centrally peaked within the core of the cluster. 
We note however that our results pertain only to regions outside the core 
radius.

We have estimated the energy input corresponding to the entropy
enhancement differently from previous works: Firstly, we have not used
any cluster scaling relations which depend on the average properties of the ICM.
 We have also used the distribution of the X-ray entropy ($K$)
with gas mass, since entropy per unit mass ($S$)
is a Lagrangian quantity. Furthermore, instead of determining an entropy floor
and then estimating an amount of energy assuming a certain density, we have
estimated the energy input from first principles.

Our result implies that the effect of energy deposition in low and high temperature
clusters is remarkably similar although the entropy enhancement processes may
differ substantially in these systems.
Secondly,
 the gas mass profiles of energy deposition per particle show
that the processes responsible for entropy enhancement in clusters affect the
gas in the central regions more than in the outer regions. Moreover, profiles for higher masses decreases faster
than those for lower masses. The basic similarity (modulo some dependence on the cluster masses) in the
profiles can provide a test for future simulations.

Taken together, our results indicate that the energy associated with
entropy enhancement is proportional to cluster mass.
Furthermore, their effect in all clusters is centrally peaked.
This suggests an energy source which must satisfy both requirements simultaneously.
As mentioned earlier, AGN feedback models satisfy the first requirement
\citep{roychowdhury04}. It is also plausible 
that the effect of the feedback is more pronounced in the inner regions, 
driving 
most of its gas outside the inner region \citep{mccarthy10, gaspari12}.

\section{Summary}
We have looked at the fractional entropy enhancement in the ICM for a sample of REXCESS clusters by comparing
the observed entropy profiles to those expected from gravitational collapse only. We first show that this entropy excess cannot be explained by only preheating plus cooling models of entropy enhancement. Since, this 
entropy excess must be sourced from non-gravitational processes, we connect this excess to any non-gravitational energy deposition in the ICM. We report, to our knowledge, the first energy deposition profiles in a large sample of clusters and also estimate the total non-gravitational energy that has been dumped into the ICM. We find that this excess energy is proportional to cluster temperature (and hence cluster mass). We show that the  entropy enhancement process in the ICM is centrally peaked and is relatively larger in low temperature clusters than in high temperature clusters. 
Our results support models of entropy enhancement through AGN feedback.

\section*{Acknowledgements}
The authors would like to thank Gabriel Pratt for providing the data on which this work is based. AC would like to thank RRI for hospitality.

\end{document}